\documentclass[journal=nanoletters,manuscript=letter]{achemso}

\usepackage[version=3]{mhchem} 
\usepackage[breaklinks]{hyperref}
\usepackage[all]{hypcap}

\hypersetup{
    plainpages=false,
    bookmarks=false,         
    unicode=false,          
    pdfborder={0 0 0}
    pdftoolbar=true,        
    pdfmenubar=true,        
    pdffitwindow=false,     
    pdfstartview={FitH},    
    pdftitle={LNCMI Annual Report 2012},    
    pdfauthor={Duncan Maude},     
    pdfproducer={LNCMI-CNRS-UJF-UPS-INSA}, 
    pdfkeywords={High} {Magnetic} {Fields}, 
    pdfnewwindow=true,      
    linktoc=section,
    colorlinks=true,       
    linkcolor=blue,          
    citecolor=red,        
    filecolor=magenta,      
    urlcolor=blue           
}

\author{P. Plochocka}
\email{paulina.plochocka@lncmi.cnrs.fr} \affiliation[LNCMI-CNRS, Grenoble and Toulouse]{Laboratoire National des Champs
Magn\'etiques Intenses, UPR 3228, CNRS-UJF-UPS-INSA, Grenoble and Toulouse, France}

\author{A. A. Mitioglu}
\affiliation[LNCMI-CNRS, Grenoble and Toulouse]{Laboratoire National des Champs Magn\'etiques Intenses, UPR 3228,
CNRS-UJF-UPS-INSA, Grenoble and Toulouse, France}

\author{D. K. Maude}
\affiliation[LNCMI-CNRS, Grenoble and Toulouse]{Laboratoire National des Champs Magn\'etiques Intenses, UPR 3228,
CNRS-UJF-UPS-INSA, Grenoble and Toulouse, France}

\author{G. L. J. A. Rikken }
\affiliation[LNCMI-CNRS, Grenoble and Toulouse]{Laboratoire National des Champs Magn\'etiques Intenses, UPR 3228,
CNRS-UJF-UPS-INSA, Grenoble and Toulouse, France}

\author{\'A. Granados del Aguila}
\affiliation[HFML, Radboud University Nijmegen]{High Field Magnet Laboratory, Institute of Molecules and Materials,
Radboud University Nijmegen, Toernooiveld 7, NL-6525 ED Nijmegen, The Netherlands}

\author{P. C. M. Christianen}
\affiliation[HFML, Radboud University Nijmegen]{High Field Magnet Laboratory, Institute of Molecules and Materials,
Radboud University Nijmegen, Toernooiveld 7, NL-6525 ED Nijmegen, The Netherlands}

\author{P. Kacman}
\affiliation [Institute of Physics, Warsaw]{Institute of Physics PAS Al. Lotników 32/46 PL-02-668 Warsaw, Poland}

\author{Hadas Shtrikman}
\affiliation [Weizmann Institute of Science, Rehovot]{Braun Center for Submicron Research, Weizmann Institute of
Science, Rehovot 76100, Israel}

\title[\texttt{achemso} ]
{High magnetic field reveals the nature of excitons in a single GaAs/AlAs core/shell nanowire}

\begin{document}


\begin{abstract}
Magneto-photoluminescence measurements of individual zinc-blende GaAs/AlAs core/shell nanowires are reported. At low
temperature a strong emission line at 1.507 eV is observed under low power (nW) excitation. Measurements performed in
high magnetic field allowed us to detect in this emission several lines associated with excitons bound to defect pairs.
Such lines were observed before in epitaxial GaAs of very high quality, as reported by Kunzel and Ploog. This
demonstrates that the optical quality of our GaAs/AlAs core/shell nanowires is comparable to the best GaAs layers grown
by molecular beam epitaxy. Moreover, strong free exciton emission is observed even at room temperature. The bright
optical emission of our nanowires in room temperature should open the way for numerous optoelectronic device
applications.
\end{abstract}


Single semiconductor nanowires (NWs) have recently attracted considerable interest due to their possible applications
in electronics or optoelectronics as NW transistors~\cite{Cui00}, lasers~\cite{Duan03} or photovoltaic
cells~\cite{Garnett10}. III-V semiconductors, such as GaAs, GaP, InAs, InP, AlAs, GaSb, may crystallize in the wurzite
(WZ) crystal structure when grown in the NW form~\cite{Lopez09,Moral07,Kriegner11}. This is unique to 1D NWs since in
3D, 2D or 0D structures these materials crystalize only in the zinc blende (ZB) form. Recent technological progress
makes it possible to control the crystal structure of the NW during the growth \emph{e.g.} to obtain pure
wurzite~\cite{Shtrikman09}, pure zinc blende~\cite{Shtrikman09a}or a mixed structure~\cite{Spirkoska09}. The different
crystal structures have different electronic band structures making it possible to manipulate the optical properties of
the NWs. For example, the coexistence of ZB and WZ phases in a single GaAs NW results in a type II band alignment with
electrons localized in the thin ZB segments and holes in the WZ regions. Additionally, controlling the ratio between ZB
and WZ phases allows to control the linear polarization of the emission from such a NW~\cite{Hoang09}.

The band structure, and therefore the photoluminescence (PL) emission energy, of a single NW depends on its crystal
structure. For zinc blende NWs it is well established that the emission corresponds to a good approximation to the
recombination energy of a free exciton in epitaxial GaAs which is 1.515 eV at Helium temperatures. In contrast, the
optical properties of wurzite type GaAs/AlAs NWs is still under debate; reported experimental values for the emission
energy vary between lower\cite{Heiss11,Ketterer11}, higher\cite{Jahn12,Hoang09} or the
same~\cite{Ketterer11a,Breuer11,Ahtapodov12} as the emission energy from ZB NWs. Even for ZB GaAs NWs the published
results show some scatter of the emission around the free exciton recombination energy. For example, Spirkoska and
co-workers report an emission at exactly the free exciton recombination energy of epitaxial GaAs~\cite{Spirkoska09},
while Jahn and co-workers show spectra where the emission occurs at an energy corresponding rather to the band gap of
GaAs $E^{ZB}_{G}$,~\cite{Jahn12} exhibiting an additional peak above $E^{ZB}_{G}$. Tiova and co-workers report emission
at 1.518~eV which is between the value of the free exciton emission energy and $E^{ZB}_{G}$~\cite{Titova06}. The
variation in the reported emission energy can be, for example, due to strain~\cite{Titova06} or the different
excitation power used in the experiment which can induce a shift of the line due to screening.

On the other hand, for ZB GaAs NWs with diameter of the order of $100$ nanometers, where confinement effects can safely
be neglected, one would expect the emission spectra to correspond to those observed in epitaxial GaAs. It is known that
in high quality GaAs samples the emission exhibits complex excitonic structure composed of free exciton X, and various
donor-bound and acceptor-bound excitons, free electron-neutral acceptor recombination and neutral-donor
neutral-acceptor pair recombination~\cite{Heim74}. Very high optical quality GaAs, possibly linked to the use of As$_4$
flux during MBE growth, exhibits a large number of reproducible emission lines over a narrow energy range
$1.504-1.511$eV (the so-called KP series first reported by Kunzel and Ploog)\cite{Kunzel80}. While these lines have
been linked to emission from excitons bound to defect pairs with different separations, possibly a carbon acceptor
paired with another acceptor, the exact nature of the recombination centers involved remains
controversial\cite{Kunzel80,Scott81,Briones82,Temkin83,Eaves84,Contour83,Heiblum83,Rao85,Skromme84,Skolnick85,Skolnick88,Charbonneau90}.
In epitaxial GaAs these emission lines are observed over a very small energy range with extremely sharp line width. To
date, most of the reported results of micro photoluminescence of single ZB GaAs NWs have an emission line width of the
order of $10-20$~meV ~\cite{Spirkoska09,Jahn12,Titova06}. In this paper we address the question of the excitonic
character of the recombination in pure ZB GaAs/AlAs NWs which posses a significantly higher optical quality. Our
samples exhibit very strong PL with a full width at half maximum of the emission line of the order of 3~meV. High
magnetic field allows us to resolve a large number of emission lines which we identify with recombination of excitons
bound to defect pairs with different separations, previously observed in very high quality epitaxial GaAs.

\begin{figure}
\includegraphics[width=6.5cm,clip,angle=0]{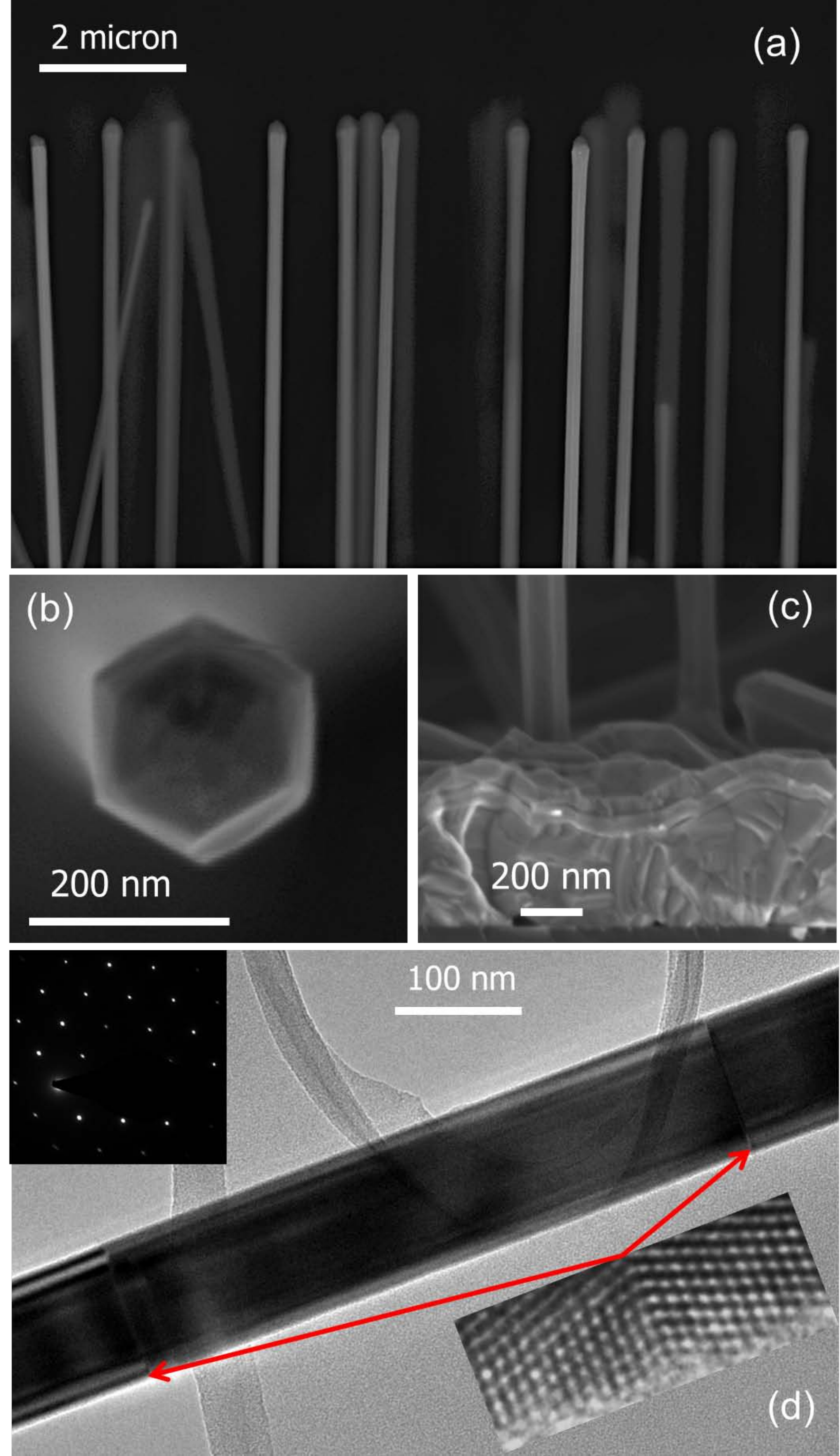}
\caption{(a) side view SEM image of the top six microns of the as grown core/shell GaAs/AlAs NWs sample depicting the
uniform aspect ratio, typical of self assisted growth.  Note the slight thickening at the tip of the wires (which are
otherwise very uniform in diameter all along ~15 micron) formed during the solidification of the gallium droplet. (b)
top view SEM image of a single core/shell GaAs/AlAs NW depicting its hexagonal cross section. (c) side view SEM image
of the bulk growth in between the NWs clearly showing the layers grown on top of the bulk GaAs during the growth of the
shell in spite of the high aspect ratio (~150) and high density of the NWs.  (d) TEM image of a single core/shell
GaAs/AlAs NW and a HR-TEM image pointing with two red arrows at rotational twin planes seen in this NW.  An electron
diffraction depicting the ZB structure can be seen in the top left corner of this image.  The images in (d) are
courtesy of Ronit Popovitz-Biro.}\label{fig:1}
\end{figure}

The GaAs/AlAs core/shell NWs were grown by molecular beam epitaxy (MBE) using the self-assisted vapor-liquid-solid
(VLS) method~\cite{Colombo08,Morral08,Krogstrup10}, on (111)-oriented Silicon bearing a native oxide layer, using Ga,
Al, and As$_4$ as source materials. After water removal at $\simeq 200^\circ$C, the Si substrate was out-gassed at high
temperature ($\simeq 600^\circ$C) in a separate chamber, before being transferred into the MBE growth chamber. The
growth was initiated by condensation of Ga at the defects in the SiO$_2$ layer and carried out at $\simeq 640^\circ$C
and a group V/III ratio of $\simeq 100$. Uniform diameter GaAs NWs were grown along the <111> direction with a high
aspect ratio ($\simeq 150$), no significant tapering and a pure zinc-blende structure, with only a few rotational twin
planes, as revealed by careful transmission electron microscopy (TEM) analysis. For growth of the uniform $\simeq 6$~nm
AlAs shell and $\simeq 12$~nm GaAs capping layer, the temperature was lowered to $\simeq 520^\circ$C.
Figure~\ref{fig:1}(a) shows a side view of the top six microns of the as grown core/shell GaAs/AlAs NWs sample
illustrating the uniform aspect ratio, typically obtained by self-assisted growth of GaAs NWs. The slightly larger
thickness at the tip of the wires forms during the solidification of the gallium droplet.  The NWs are otherwise very
uniform in diameter all along $\sim 15$ microns. A top view scanning electron microscope (SEM) image of a single
core/shell GaAs/AlAs can be seen in Figure~\ref{fig:1}(b) exhibiting a clear hexagonal cross section. At such high
aspect ratio and high NWs density, one has to bare in mind the possibility of shadowing. We have excluded shadowing
here by looking at the bulk growth in between the NWs, where the layers grown during the growth of the shell can
clearly be seen by SEM (Figure~\ref{fig:1}(c)) in spite of the high aspect ratio ($\sim 150$) and the high density of
NWs. Figure~\ref{fig:1}(d) shows a TEM image of a single core/shell GaAs/AlAs NW and a HR-TEM image of a single
rotational twin plane occasionally found along such NWs. The electron diffraction (inset) confirms the existence of a
ZB structure.

\begin{figure}[h]
\includegraphics[width=8.5cm,clip,angle=0]{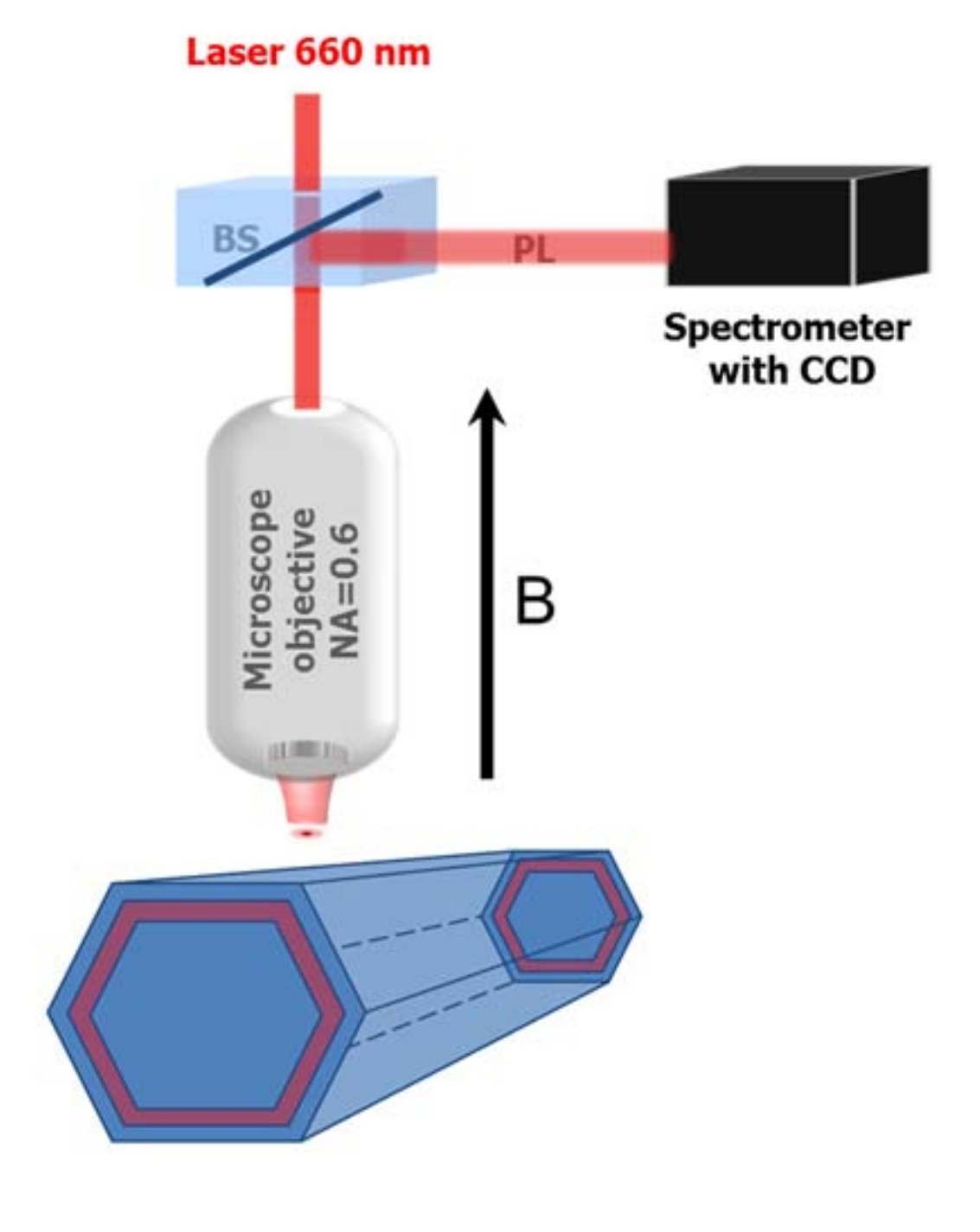}
\caption{Schematic of the micro-photoluminescence setup used to investigate excitons in hexagonal GaAs/AlAs core/shell
NWs.}\label{fig:2}
\end{figure}

For the optical measurements, the NWs were dispersed onto a Si substrate with their <111> growth axis parallel to the
surface. The average separation between wires is $d \approx 30 \mu$m to facilitate the investigation of the
photoluminescence from individual NWs.  The sample was placed in a micro-photoluminescence ($\mu$-PL) setup containing
piezoelectric $x-y-z$ translation stages and a microscope objective (see Figure~\ref{fig:2}). The $\mu$-PL system was
kept at a temperature of $1.8$ K in a cryostat placed in a resistive magnet producing a magnetic field of up to $23.5$
T. The field was applied in the Faraday configuration, perpendicular to the NW <111> growth axis. The PL of the NWs was
excited with a diode laser at $660$ nm. Both the exciting and the collected light were transmitted through a monomode
fiber coupled directly to the microscope objective. The diameter of the excitation beam was $\simeq 1\mu m$. The
emission from the sample was dispersed in a spectrometer equipped with a CCD camera. The polarization resolved
measurements were performed using free beam optics and a microscope objective. Several NWs have been investigated and
all give almost identical results. In this letter we show representative results from our acquired data set.

\begin{figure}[t]
\includegraphics[width=8.5cm,clip,angle=0]{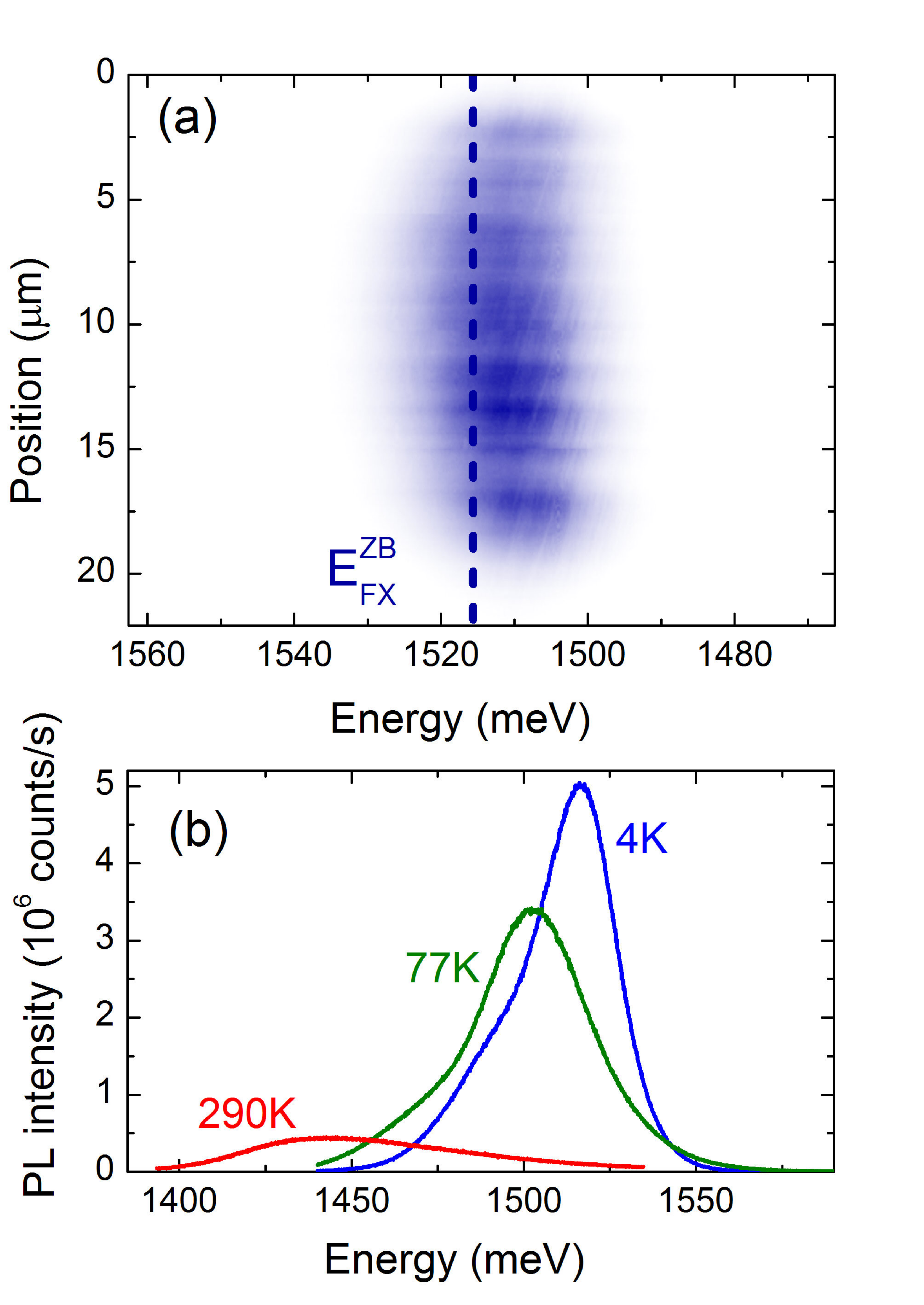}
\caption{(a) Color plot showing typical $\mu$-PL spectra measured at $T=4$~K as a function of position along a single
GaAs/AlAs core/shell NM. The dashed blue line indicates the energy of the emission of a free exciton in epitaxial GaAs.
(b) Temperature dependence of the PL of the free exciton line of the NW under excitation power $\simeq 11 \mu$W
($\simeq 1100$~W/cm$^2$). Note that even at room temperature the PL emission saturates the CCD for integrations times
$\geq 100$~ms.}\label{fig:3}
\end{figure}

First we analyze the data obtained without magnetic field in order to characterize the investigated NWs. In
Figure~~\ref{fig:3}(a) a color plot of the PL spectra as a function of position along the NW is presented. Under the
excitation conditions (power) used the observed emission energy is dominated by emission close to the energy of the
free exciton recombination in GaAs (indicated in the figure by the solid blue line). Importantly, the spectra do not
change as a function of position along the NW which confirms that it posses a pure ZB
structure~\cite{Spirkoska09,Jahn12}. In Figure~~\ref{fig:3}(b) we show the temperature dependence of the free exciton
PL line. Here bright PL is observed even at room temperature, when we excite \emph{a small portion of a single NW}
($\simeq 1 \mu$m out of a total length $\simeq 15 \mu$m \emph{i.e.} only 5\% of the NW). The emission, with around $5
\times 10^5$~counts per second, saturates the CCD for integration times $\geq 100$~ms. Strong room temperature emission
has previously been reported for macro-PL on a large number of ZB GaAs NWs with slightly different diameters and
different lengths~\cite{Dhaka12}.

\begin{figure}[h!]
\includegraphics[width=8.5cm,clip,angle=0]{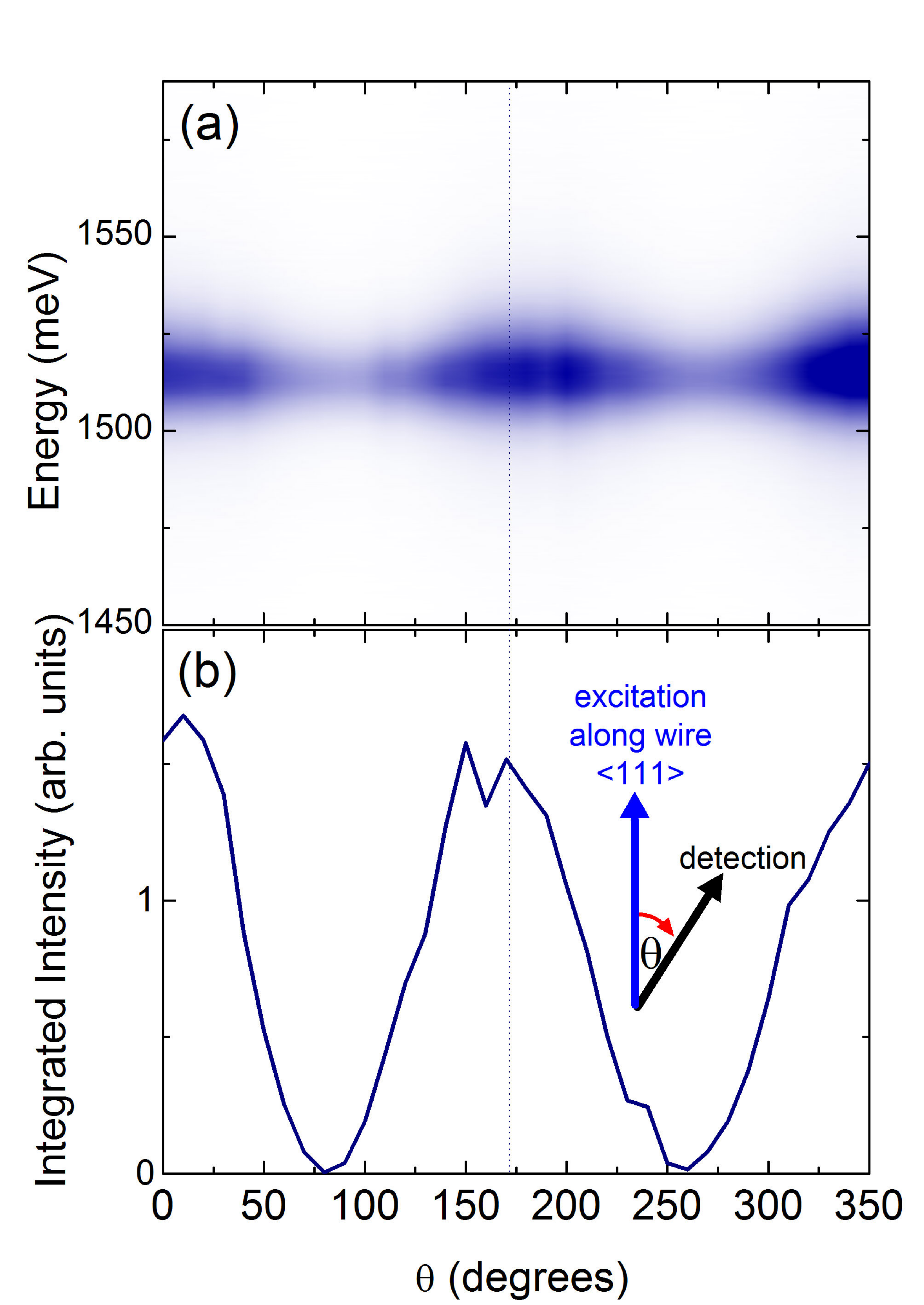}
\caption{(a) PL emission spectra as a function of angle ($\theta$) between the linear polarization of the excitation
and collection. The polarization of the excitation is always aligned along the NW.  (b) Integrated intensity of the PL
as a function of the angle between the linear polarization of the excitation and collection.}\label{fig:4}
\end{figure}

Linear polarization resolved measurements can also be used to differentiate between ZB and WZ structure. It is well
established that for a pure ZB crystal structure for excitation and detection along the NW the free exciton emission is
much stronger than for excitation and detection across the NW~\cite{Ruda05,Titova06}. A strong polarization anisotropy
was also observed in InP NWs~\cite{Wang01} and is the result of the dielectric mismatch between the NW and its
surrounding, which leads to the suppression of the component of the electric field perpendicular to the
NW~\cite{Ruda05}. This is not the case for NWs with WZ crystal structure, where the c-axis of the crystal is oriented
along the long axis of the NW and the optical selection rules require that the free exciton emission is allowed only if
the dielectric dipole moment is perpendicular to the c-axis of the crystal. This results in emission from WZ NWs which
is strongly polarized across the NW~\cite{Mishra07,Birman59}.

The measured linear polarization of the free exciton emission from the NW is presented in Figure~~\ref{fig:4}(a-b). A
linearly polarized excitation was used with the polarizer always aligned along the NW which corresponds to maximum PL
emission. The linear polarization was analyzed by rotating the linear polarizer for detection with respect to the
excitation polarizer. In Figure~~\ref{fig:4}(a-b) the emission spectra as a function of angle are presented together
with the integrated intensity. We observe minima in the amplitude of the collected emission for the configuration in
which the polarization of the emission is orthogonal (crossed) with the polarization of the excitation, and maxima when
the polarizers are parallel. Thus the polarization resolved measurements confirm that the investigated NW consists of a
pure ZB structure.

\begin{figure}[t]
\includegraphics[width=8.5cm,clip,angle=0]{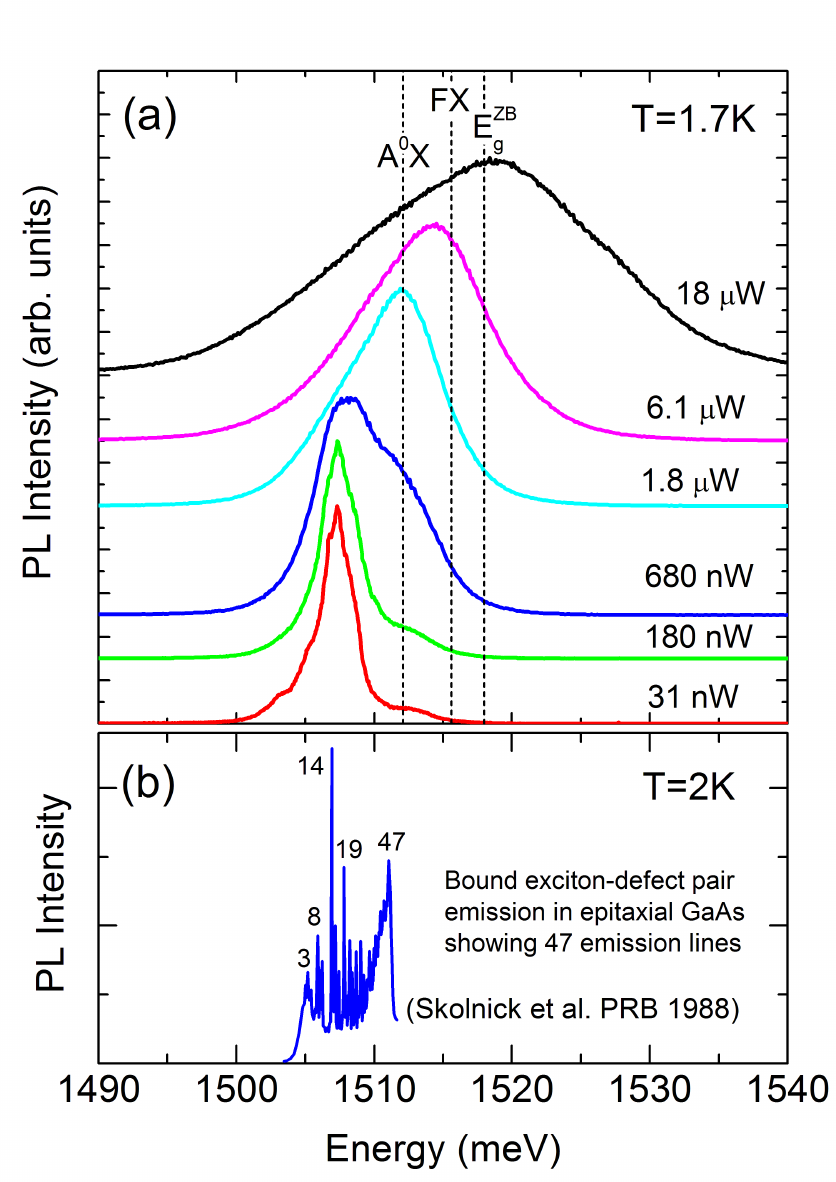}
\caption{(a) Typical $\mu$-PL spectra of the single GaAs NW measured for different excitation powers (3.1 -
1800~W/cm$^2$ assuming a spot size of $1\mu$m$^2$). The spectra have been normalized and shifted vertically for
clarity. The data have been taken at B$=0$T and $1.7$K. (b) PL of high quality MBE grown GaAs is shown for comparison
(taken from Skolnick \emph{et al.}\cite{Skolnick88}).}\label{fig:5}
\end{figure}

Typical $\mu$-PL spectra of a single GaAs/AlAs core/shell NW measured for different excitation power are presented in
Figure~\ref{fig:5}(a). The broken vertical lines indicate the recombination energy of an exciton bound to a neutral
acceptor (A$^0$X), a free exciton (FX) and the value of the epitaxial ZB GaAs band gap (E$_g^{ZB}$). In
Figure~\ref{fig:5}(b) we show for comparison PL from high quality MBE grown GaAs measured by Skolnick and
co-workers\cite{Skolnick85,Skolnick88}. Forty seven closely spaced lines are resolved in the original data (not all are
visible in Figure\ref{fig:5}(b)), the energetic position of the lines is reproducible between different samples, and
the strong line at 1.5069~eV (labeled 14) dominates the emission in all samples
investigated\cite{Skolnick85,Skolnick88}. These lines were identified with emission from excitons bound to defect pairs
with different separations. Confinement effects can be neglected for our $\simeq 100$~nm diameter (core) NWs so that
the PL results can be directly compared. We focus for the moment on the PL taken at the lowest power for which two
lines are observed. The strong line at $\simeq 1.507$~eV which we attribute to broadened emission from excitons bound
to defect pairs (the energy corresponds almost exactly to the strong line 14 in the KP series). Indeed, upon closer
inspection we find that this line consists of a double peak with a number of shoulders suggesting that it is made up of
a large number of broadened lines. The weaker line around 1.514 eV can be associated with the recombination of an
exciton bound to a neutral acceptor ($A^{0}-X$). With increasing power the amplitude of the $A^{0}-X$ emission
increases with respect to the bound exciton-defect pair emission with an increasing contribution from free exciton
emission at higher powers. Further increase in the power induces a red shift of the free exciton emission towards the
GaAs band gap energy presumably due to the screening of the exciton binding energy at high carrier densities. The line
at $\simeq 1.507$~eV associated with excitons bound to defect pairs was not observed in previously published data,
possibly due to either the excitation power used or the larger linewidths\cite{Spirkoska09,Jahn12}.

\begin{figure}[]
\includegraphics[width=17cm,clip,angle=0]{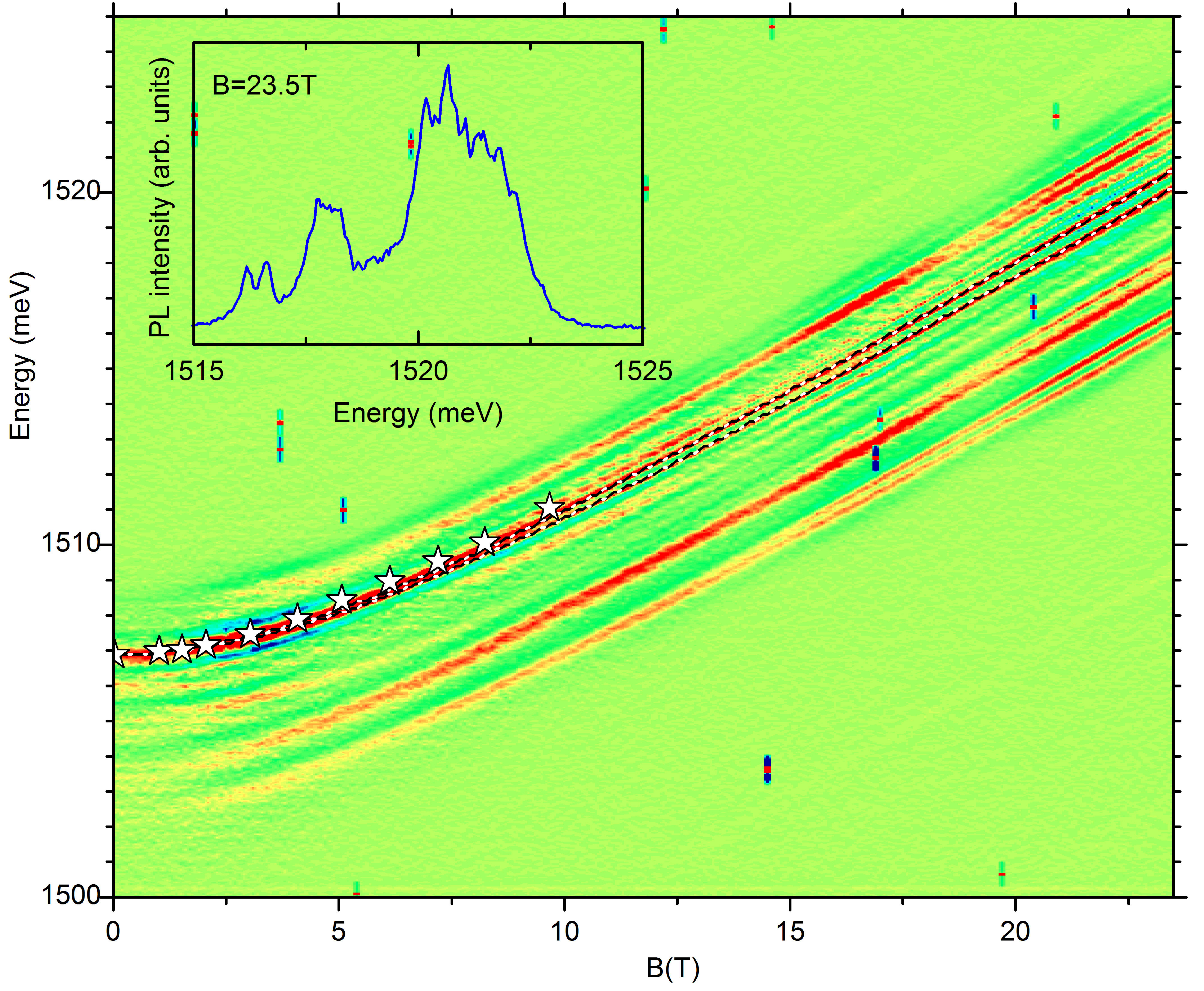}
\caption{Color plot of the differential $\mu$PL spectra showing the evolution of the emission as a function of magnetic
field applied perpendicular to the NW at $T=1.7$K. Several sharp emission lines are resolved which show a large
diamagnetic shift and Zeeman splitting at high magnetic fields.  The white symbols show for comparison the shift of the
strong line (14) of the bound exciton-defect pair recombination in epitaxial GaAs of Skolnick \emph{et
al.}.\cite{Skolnick88} The broken lines are a fit to the diamagnetic shift as described in the text. The inset shows a
$\mu$PL spectra taken at $B=23.5$~T}\label{fig:6}
\end{figure}

Further support for our assignment of the $\simeq 1.507$~eV line to bound exciton-defect pair recombination is provided
by the $\mu$PL data in magnetic fields as high as $23.5$~T. The spectra were acquired at $T=1.7$~K using low power
excitation under which the line at 1.5069~eV dominates the zero field spectrum. In Figure~\ref{fig:6} we show a color
plot of the differential PL obtained by subtracting a suitably averaged PL spectra\cite{DifferentialPL12}. With
increasing magnetic field the intensity of the emission increases dramatically, accompanied by a reduction in the line
width. Several emission lines can be seen that exhibit a large diamagnetic shift ($\propto B^2$) at low fields, which
becomes linear in high magnetic fields. The observed energetic position and shift with magnetic field corresponds
exactly to that of the strong line 14 of the bound exciton-defect pair recombination reported by Skolnick and
co-workers,\cite{Skolnick88} in high quality MBE grown GaAs, indicated by the symbols in Figure~\ref{fig:6}.

\begin{figure}[h]
\includegraphics[width=8.5cm,clip,angle=0]{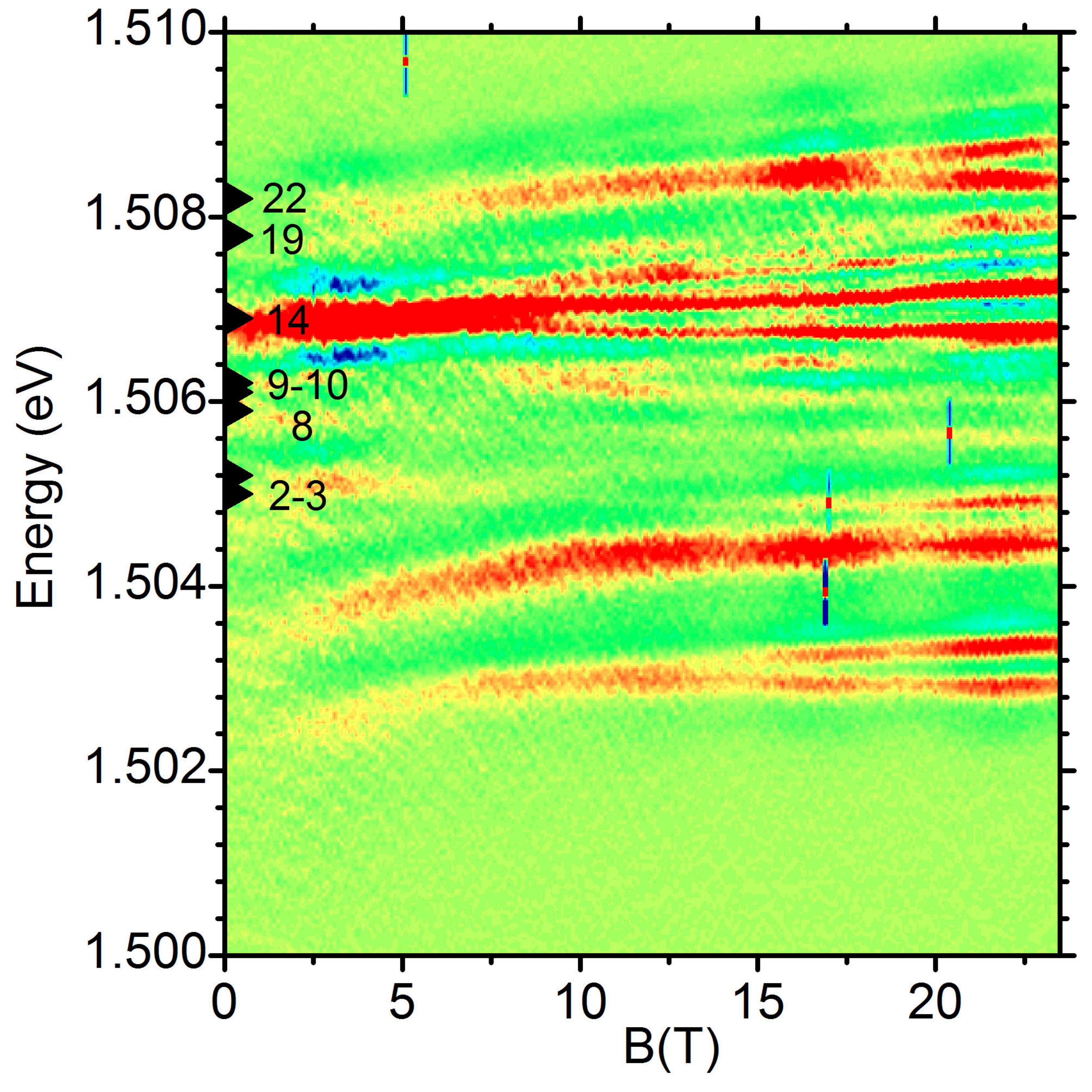}
\caption{Color plot of the of the differential $\mu$PL spectra as a function of magnetic field after subtraction of the
large diamagnetic shift. Strong features observed in the bound exciton-defect pair spectrum in GaAs are indicated by
white arrow heads and labeled as in Skolnick \emph{et al.}\cite{Skolnick88}}\label{fig:7}
\end{figure}

For $B \geq 10$~T a Zeeman splitting of the lines is observed. This can be seen in more detail in Figure~\ref{fig:7}
where we have subtracted the large diamagnetic shift, by fitting the expression for the ground state of a two
dimensional harmonic oscillator in perpendicular magnetic field, $E=\hbar \sqrt{\omega_0^2 + (\omega_c / 2)^2}$, to the
strongest line at $1.5069$~eV at $B=0$~T. Here $\omega_0$ is the harmonic trap frequency and $\omega_c = eB/m^*$ is the
cyclotron frequency. An excellent fit (dashed black lines in Figure~\ref{fig:6}) is achieved  using $\hbar \omega_0
\simeq 5.6$~meV and $m^*\simeq 0.0745 m_e$, which is reasonable for an effective mass electron weakly bound in a
Coulomb potential. All of the observed emission lines have approximately the same slope at high magnetic fields
demonstrating that the excitons involved, all have an orbital quantum number $N=0$. The Zeeman splitting has been
reproduced using an effective g-factor $\mid g \mid \approx0.19$ for an assumed effective spin of $1/2$.

Strong features observed in the bound exciton-defect pair spectrum in GaAs are indicated by white arrow heads in
Figure~\ref{fig:7} and labeled using the notation of Skolnick and co-workers\cite{Skolnick88}. In Figure~\ref{fig:7} at
high magnetic fields some lines do not extrapolate correctly to zero field because their diamagnetic shift is slightly
different from that of the 1.5069~eV line used for the correction. It should be noted that subtracting the diamagnetic
shift has no effect on the energetic position of the lines at $B=0$, which should be used for the comparison. Clearly
our data reproduces quite nicely the prominent features of the bound exciton-defect pair spectrum of epitaxial GaAs.

In GaAs the Zeeman splitting of the peaks is well described using an effective spin Hamiltonian for an $S=1/2$ electron
and a $J=3/2$ hole in orthorhombic symmetry involving isotropic and anisotropic hole g-factors together with crystal
field parameters which can give rise to a zero field splitting\cite{Skolnick88}. In our measurements the magnetic field
is applied perpendicular to the NW \emph{i.e.} perpendicular to the <111> growth direction. However, the orientation of
the magnetic field within the (111) plane is unknown so that we are unable to compare our results with the detailed
predictions of the effective spin Hamiltonian.

To conclude, we have observed emission lines in the photoluminescence of single core/shell GaAs/AlAs NWs with a pure
zinc-blende structure in the energy range $1.504-1.511$eV. The magnetic field dependence of the emission allows us to
unambiguously identify the emission with the recombination of excitons bound to defect pairs with different
separations, previously observed only in very high quality MBE grown GaAs. This, together with the strong PL emission
at room temperature demonstrates the very high optical quality of our GaAs NWs. The NWs, which can easily be dispersed
onto a Si substrate, should open the way for numerous optoelectronic applications based on the well established Si
device technology.

\acknowledgement

We would like to thank G. Kopnov for some help with the early stages of sample preparation, Ronit Popovitz-Biro for the
TEM mesurements and I. Breslavetz for assistance with the $\mu$PL measurements. Part of this work was supported by
EUROMAGNET II under the SP7 transnational access program of the European Union under contract number 228043. We would
also like to acknowledge partial support by the Israeli Science Foundation grant \#530/08 and Israeli Ministry of
Science grant \#3-6799.


\begin{mcitethebibliography}{40}
\providecommand*\natexlab[1]{#1} \providecommand*\mciteSetBstSublistMode[1]{}
\providecommand*\mciteSetBstMaxWidthForm[2]{} \providecommand*\mciteBstWouldAddEndPuncttrue
  {\def\EndOfBibitem{\unskip.}}
\providecommand*\mciteBstWouldAddEndPunctfalse
  {\let\EndOfBibitem\relax}
\providecommand*\mciteSetBstMidEndSepPunct[3]{} \providecommand*\mciteSetBstSublistLabelBeginEnd[3]{}
\providecommand*\EndOfBibitem{} \mciteSetBstSublistMode{f}
\mciteSetBstMaxWidthForm{subitem}{(\alph{mcitesubitemcount})} \mciteSetBstSublistLabelBeginEnd
  {\mcitemaxwidthsubitemform\space}
  {\relax}
  {\relax}

\bibitem[Cui et~al.(2000)Cui, Duan, Hu, and Lieber]{Cui00}
Cui,~Y.; Duan,~X.; Hu,~J.; Lieber,~C.~M. \emph{The Journal of Physical
  Chemistry B} \textbf{2000}, \emph{104}, 5213--5216\relax
\mciteBstWouldAddEndPuncttrue \mciteSetBstMidEndSepPunct{\mcitedefaultmidpunct}
{\mcitedefaultendpunct}{\mcitedefaultseppunct}\relax \EndOfBibitem
\bibitem[Duan et~al.(2003)Duan, Huang, Agarwal, and Lieber]{Duan03}
Duan,~X.; Huang,~Y.; Agarwal,~R.; Lieber,~C.~M. \emph{Nature} \textbf{2003},
  \emph{421}, 241--245\relax
\mciteBstWouldAddEndPuncttrue \mciteSetBstMidEndSepPunct{\mcitedefaultmidpunct}
{\mcitedefaultendpunct}{\mcitedefaultseppunct}\relax \EndOfBibitem
\bibitem[Garnett and Yang(2010)Garnett, and Yang]{Garnett10}
Garnett,~E.; Yang,~P. \emph{Nano Letters} \textbf{2010}, \emph{10}, 1082--1087,
  PMID: 20108969\relax
\mciteBstWouldAddEndPuncttrue \mciteSetBstMidEndSepPunct{\mcitedefaultmidpunct}
{\mcitedefaultendpunct}{\mcitedefaultseppunct}\relax \EndOfBibitem
\bibitem[Lopez et~al.(2009)Lopez, Hemesath, and Lauhon]{Lopez09}
Lopez,~F.~J.; Hemesath,~E.~R.; Lauhon,~L.~J. \emph{Nano Letters} \textbf{2009},
  \emph{9}, 2774--2779, PMID: 19527044\relax
\mciteBstWouldAddEndPuncttrue \mciteSetBstMidEndSepPunct{\mcitedefaultmidpunct}
{\mcitedefaultendpunct}{\mcitedefaultseppunct}\relax \EndOfBibitem
\bibitem[Fontcuberta~i Morral et~al.(2007)Fontcuberta~i Morral, Arbiol, Prades,
  Cirera, and Morante]{Moral07}
Fontcuberta~i Morral,~A.; Arbiol,~J.; Prades,~J.; Cirera,~A.; Morante,~J.
  \emph{Advanced Materials} \textbf{2007}, \emph{19}, 1347--1351\relax
\mciteBstWouldAddEndPuncttrue \mciteSetBstMidEndSepPunct{\mcitedefaultmidpunct}
{\mcitedefaultendpunct}{\mcitedefaultseppunct}\relax \EndOfBibitem
\bibitem[Kriegner et~al.(2011)Kriegner, Panse, Mandl, Dick, Keplinger, Persson,
  Caroff, Ercolani, Sorba, Bechstedt, Stangl, and Bauer]{Kriegner11}
Kriegner,~D.; Panse,~C.; Mandl,~B.; Dick,~K.~A.; Keplinger,~M.; Persson,~J.~M.;
  Caroff,~P.; Ercolani,~D.; Sorba,~L.; Bechstedt,~F.; Stangl,~J.; Bauer,~G.
  \emph{Nano Letters} \textbf{2011}, \emph{11}, 1483--1489\relax
\mciteBstWouldAddEndPuncttrue \mciteSetBstMidEndSepPunct{\mcitedefaultmidpunct}
{\mcitedefaultendpunct}{\mcitedefaultseppunct}\relax \EndOfBibitem
\bibitem[Shtrikman et~al.(2009)Shtrikman, Popovitz-Biro, Kretinin, Houben,
  Heiblum, Bukala, Galicka, Buczko, and Kacman]{Shtrikman09}
Shtrikman,~H.; Popovitz-Biro,~R.; Kretinin,~A.; Houben,~L.; Heiblum,~M.;
  Bukala,~M.; Galicka,~M.; Buczko,~R.; Kacman,~P. \emph{Nano Letters}
  \textbf{2009}, \emph{9}, 1506--1510, PMID: 19253998\relax
\mciteBstWouldAddEndPuncttrue \mciteSetBstMidEndSepPunct{\mcitedefaultmidpunct}
{\mcitedefaultendpunct}{\mcitedefaultseppunct}\relax \EndOfBibitem
\bibitem[Shtrikman et~al.(2009)Shtrikman, Popovitz-Biro, Kretinin, and
  Heiblum]{Shtrikman09a}
Shtrikman,~H.; Popovitz-Biro,~R.; Kretinin,~A.; Heiblum,~M. \emph{Nano Letters}
  \textbf{2009}, \emph{9}, 215--219\relax
\mciteBstWouldAddEndPuncttrue \mciteSetBstMidEndSepPunct{\mcitedefaultmidpunct}
{\mcitedefaultendpunct}{\mcitedefaultseppunct}\relax \EndOfBibitem
\bibitem[Spirkoska et~al.(2009)Spirkoska, Arbiol, Gustafsson, Conesa-Boj, Glas,
  Zardo, Heigoldt, Gass, Bleloch, Estrade, Kaniber, Rossler, Peiro, Morante,
  Abstreiter, Samuelson, and Fontcuberta~i Morral]{Spirkoska09}
Spirkoska,~D. et~al.  \emph{Phys. Rev. B} \textbf{2009}, \emph{80},
  245325\relax
\mciteBstWouldAddEndPuncttrue \mciteSetBstMidEndSepPunct{\mcitedefaultmidpunct}
{\mcitedefaultendpunct}{\mcitedefaultseppunct}\relax \EndOfBibitem
\bibitem[Hoang et~al.(2009)Hoang, Moses, Zhou, Dheeraj, Fimland, and
  Weman]{Hoang09}
Hoang,~T.~B.; Moses,~A.~F.; Zhou,~H.~L.; Dheeraj,~D.~L.; Fimland,~B.~O.;
  Weman,~H. \emph{Applied Physics Letters} \textbf{2009}, \emph{94},
  133105\relax
\mciteBstWouldAddEndPuncttrue \mciteSetBstMidEndSepPunct{\mcitedefaultmidpunct}
{\mcitedefaultendpunct}{\mcitedefaultseppunct}\relax \EndOfBibitem
\bibitem[Heiss et~al.(2011)Heiss, Conesa-Boj, Ren, Tseng, Gali, Rudolph,
  Uccelli, Peir\'o, Morante, Schuh, Reiger, Kaxiras, Arbiol, and Fontcuberta~i
  Morral]{Heiss11}
Heiss,~M.; Conesa-Boj,~S.; Ren,~J.; Tseng,~H.-H.; Gali,~A.; Rudolph,~A.;
  Uccelli,~E.; Peir\'o,~F.; Morante,~J.~R.; Schuh,~D.; Reiger,~E.; Kaxiras,~E.;
  Arbiol,~J.; Fontcuberta~i Morral,~A. \emph{Phys. Rev. B} \textbf{2011},
  \emph{83}, 045303\relax
\mciteBstWouldAddEndPuncttrue \mciteSetBstMidEndSepPunct{\mcitedefaultmidpunct}
{\mcitedefaultendpunct}{\mcitedefaultseppunct}\relax \EndOfBibitem
\bibitem[Ketterer et~al.(2011)Ketterer, Heiss, Livrozet, Rudolph, Reiger, and
  Fontcuberta~i Morral]{Ketterer11}
Ketterer,~B.; Heiss,~M.; Livrozet,~M.~J.; Rudolph,~A.; Reiger,~E.;
  Fontcuberta~i Morral,~A. \emph{Phys. Rev. B} \textbf{2011}, \emph{83},
  125307\relax
\mciteBstWouldAddEndPuncttrue \mciteSetBstMidEndSepPunct{\mcitedefaultmidpunct}
{\mcitedefaultendpunct}{\mcitedefaultseppunct}\relax \EndOfBibitem
\bibitem[Jahn et~al.(2012)Jahn, L\"ahnemann, Pf\"uller, Brandt, Breuer,
  Jenichen, Ramsteiner, Geelhaar, and Riechert]{Jahn12}
Jahn,~U.; L\"ahnemann,~J.; Pf\"uller,~C.; Brandt,~O.; Breuer,~S.; Jenichen,~B.;
  Ramsteiner,~M.; Geelhaar,~L.; Riechert,~H. \emph{Phys. Rev. B} \textbf{2012},
  \emph{85}, 045323\relax
\mciteBstWouldAddEndPuncttrue \mciteSetBstMidEndSepPunct{\mcitedefaultmidpunct}
{\mcitedefaultendpunct}{\mcitedefaultseppunct}\relax \EndOfBibitem
\bibitem[Ketterer et~al.(2011)Ketterer, Heiss, Uccelli, Arbiol, and
  Fontcuberta~i Morral]{Ketterer11a}
Ketterer,~B.; Heiss,~M.; Uccelli,~E.; Arbiol,~J.; Fontcuberta~i Morral,~A.
  \emph{ACS Nano} \textbf{2011}, \emph{5}, 7585--7592\relax
\mciteBstWouldAddEndPuncttrue \mciteSetBstMidEndSepPunct{\mcitedefaultmidpunct}
{\mcitedefaultendpunct}{\mcitedefaultseppunct}\relax \EndOfBibitem
\bibitem[Breuer et~al.(2011)Breuer, Pfuller, Flissikowski, Brandt, Grahn,
  Geelhaar, and Riechert]{Breuer11}
Breuer,~S.; Pfuller,~C.; Flissikowski,~T.; Brandt,~O.; Grahn,~H.~T.;
  Geelhaar,~L.; Riechert,~H. \emph{Nano Letters} \textbf{2011}, \emph{11},
  1276--1279\relax
\mciteBstWouldAddEndPuncttrue \mciteSetBstMidEndSepPunct{\mcitedefaultmidpunct}
{\mcitedefaultendpunct}{\mcitedefaultseppunct}\relax \EndOfBibitem
\bibitem[Ahtapodov et~al.(2012)Ahtapodov, Todorovic, Olk, Mjaland, Slattnes,
  Dheeraj, van Helvoort, Fimland, and Weman]{Ahtapodov12}
Ahtapodov,~L.; Todorovic,~J.; Olk,~P.; Mjaland,~T.; Slattnes,~P.;
  Dheeraj,~D.~L.; van Helvoort,~A. T.~J.; Fimland,~B.-O.; Weman,~H. \emph{Nano
  Letters} \textbf{2012}, \emph{12}, 6090--6095\relax
\mciteBstWouldAddEndPuncttrue \mciteSetBstMidEndSepPunct{\mcitedefaultmidpunct}
{\mcitedefaultendpunct}{\mcitedefaultseppunct}\relax \EndOfBibitem
\bibitem[Titova et~al.(2006)Titova, Hoang, Jackson, Smith, Yarrison-Rice, Kim,
  Joyce, Tan, and Jagadish]{Titova06}
Titova,~L.~V.; Hoang,~T.~B.; Jackson,~H.~E.; Smith,~L.~M.;
  Yarrison-Rice,~J.~M.; Kim,~Y.; Joyce,~H.~J.; Tan,~H.~H.; Jagadish,~C.
  \emph{Applied Physics Letters} \textbf{2006}, \emph{89}, 173126\relax
\mciteBstWouldAddEndPuncttrue \mciteSetBstMidEndSepPunct{\mcitedefaultmidpunct}
{\mcitedefaultendpunct}{\mcitedefaultseppunct}\relax \EndOfBibitem
\bibitem[Heim and Hiesinger(1974)Heim, and Hiesinger]{Heim74}
Heim,~U.; Hiesinger,~P. \emph{Phys. Stat. Sol. (b)} \textbf{1974}, \emph{66},
  461\relax
\mciteBstWouldAddEndPuncttrue \mciteSetBstMidEndSepPunct{\mcitedefaultmidpunct}
{\mcitedefaultendpunct}{\mcitedefaultseppunct}\relax \EndOfBibitem
\bibitem[Kunzel and Ploog(1980)Kunzel, and Ploog]{Kunzel80}
Kunzel,~H.; Ploog,~K. \emph{Applied Physics Letters} \textbf{1980}, \emph{37},
  416--418\relax
\mciteBstWouldAddEndPuncttrue \mciteSetBstMidEndSepPunct{\mcitedefaultmidpunct}
{\mcitedefaultendpunct}{\mcitedefaultseppunct}\relax \EndOfBibitem
\bibitem[Scott et~al.(1981)Scott, Duggan, Dawson, and Weimann]{Scott81}
Scott,~G.~B.; Duggan,~G.; Dawson,~P.; Weimann,~G. \emph{Journal of Applied
  Physics} \textbf{1981}, \emph{52}, 6888--6894\relax
\mciteBstWouldAddEndPuncttrue \mciteSetBstMidEndSepPunct{\mcitedefaultmidpunct}
{\mcitedefaultendpunct}{\mcitedefaultseppunct}\relax \EndOfBibitem
\bibitem[Briones and Collins(1982)Briones, and Collins]{Briones82}
Briones,~F.; Collins,~D. \emph{Journal of Electronic Materials} \textbf{1982},
  \emph{11}, 847--866\relax
\mciteBstWouldAddEndPuncttrue \mciteSetBstMidEndSepPunct{\mcitedefaultmidpunct}
{\mcitedefaultendpunct}{\mcitedefaultseppunct}\relax \EndOfBibitem
\bibitem[Temkin and Hwang(1983)Temkin, and Hwang]{Temkin83}
Temkin,~H.; Hwang,~J. C.~M. \emph{Applied Physics Letters} \textbf{1983},
  \emph{42}, 178--180\relax
\mciteBstWouldAddEndPuncttrue \mciteSetBstMidEndSepPunct{\mcitedefaultmidpunct}
{\mcitedefaultendpunct}{\mcitedefaultseppunct}\relax \EndOfBibitem
\bibitem[Eaves and Halliday(1984)Eaves, and Halliday]{Eaves84}
Eaves,~L.; Halliday,~D.~P. \emph{Journal of Physics C: Solid State Physics}
  \textbf{1984}, \emph{17}, L705--L709\relax
\mciteBstWouldAddEndPuncttrue \mciteSetBstMidEndSepPunct{\mcitedefaultmidpunct}
{\mcitedefaultendpunct}{\mcitedefaultseppunct}\relax \EndOfBibitem
\bibitem[Contour et~al.(1983)Contour, Neu, Leroux, Chaix, Levesque, and
  Etienne]{Contour83}
Contour,~J.~P.; Neu,~G.; Leroux,~M.; Chaix,~C.; Levesque,~B.; Etienne,~P.
  \emph{Journal of Vacuum Science \& Technology B: Microelectronics and
  Nanometer Structures} \textbf{1983}, \emph{1}, 811--815\relax
\mciteBstWouldAddEndPuncttrue \mciteSetBstMidEndSepPunct{\mcitedefaultmidpunct}
{\mcitedefaultendpunct}{\mcitedefaultseppunct}\relax \EndOfBibitem
\bibitem[Heiblum et~al.(1983)Heiblum, Mendez, and Osterling]{Heiblum83}
Heiblum,~M.; Mendez,~E.~E.; Osterling,~L. \emph{Journal of Applied Physics}
  \textbf{1983}, \emph{54}, 6982--6988\relax
\mciteBstWouldAddEndPuncttrue \mciteSetBstMidEndSepPunct{\mcitedefaultmidpunct}
{\mcitedefaultendpunct}{\mcitedefaultseppunct}\relax \EndOfBibitem
\bibitem[Rao et~al.(1985)Rao, Alexandre, Masson, Allovon, and Goldstein]{Rao85}
Rao,~E. V.~K.; Alexandre,~F.; Masson,~J.~M.; Allovon,~M.; Goldstein,~L.
  \emph{Journal of Applied Physics} \textbf{1985}, \emph{57}, 503--508\relax
\mciteBstWouldAddEndPuncttrue \mciteSetBstMidEndSepPunct{\mcitedefaultmidpunct}
{\mcitedefaultendpunct}{\mcitedefaultseppunct}\relax \EndOfBibitem
\bibitem[Skromme et~al.(1984)Skromme, Stillman, Calawa, and Metze]{Skromme84}
Skromme,~B.~J.; Stillman,~G.~E.; Calawa,~A.~R.; Metze,~G.~M. \emph{Applied
  Physics Letters} \textbf{1984}, \emph{44}, 240--242\relax
\mciteBstWouldAddEndPuncttrue \mciteSetBstMidEndSepPunct{\mcitedefaultmidpunct}
{\mcitedefaultendpunct}{\mcitedefaultseppunct}\relax \EndOfBibitem
\bibitem[Skolnick et~al.(1985)Skolnick, Harris, Tu, Brennan, and
  Sturge]{Skolnick85}
Skolnick,~M.~S.; Harris,~T.~D.; Tu,~C.~W.; Brennan,~T.~M.; Sturge,~M.~D.
  \emph{Applied Physics Letters} \textbf{1985}, \emph{46}, 427--429\relax
\mciteBstWouldAddEndPuncttrue \mciteSetBstMidEndSepPunct{\mcitedefaultmidpunct}
{\mcitedefaultendpunct}{\mcitedefaultseppunct}\relax \EndOfBibitem
\bibitem[Skolnick et~al.(1988)Skolnick, Halliday, and Tu]{Skolnick88}
Skolnick,~M.~S.; Halliday,~D.~P.; Tu,~C.~W. \emph{Phys. Rev. B} \textbf{1988},
  \emph{38}, 4165--4179\relax
\mciteBstWouldAddEndPuncttrue \mciteSetBstMidEndSepPunct{\mcitedefaultmidpunct}
{\mcitedefaultendpunct}{\mcitedefaultseppunct}\relax \EndOfBibitem
\bibitem[Charbonneau and Thewalt(1990)Charbonneau, and Thewalt]{Charbonneau90}
Charbonneau,~S.; Thewalt,~M. L.~W. \emph{Phys. Rev. B} \textbf{1990},
  \emph{41}, 8221--8228\relax
\mciteBstWouldAddEndPuncttrue \mciteSetBstMidEndSepPunct{\mcitedefaultmidpunct}
{\mcitedefaultendpunct}{\mcitedefaultseppunct}\relax \EndOfBibitem
\bibitem[Colombo et~al.(2008)Colombo, Spirkoska, Frimmer, Abstreiter, and
  Fontcuberta~i Morral]{Colombo08}
Colombo,~C.; Spirkoska,~D.; Frimmer,~M.; Abstreiter,~G.; Fontcuberta~i
  Morral,~A. \emph{Phys. Rev. B} \textbf{2008}, \emph{77}, 155326\relax
\mciteBstWouldAddEndPuncttrue \mciteSetBstMidEndSepPunct{\mcitedefaultmidpunct}
{\mcitedefaultendpunct}{\mcitedefaultseppunct}\relax \EndOfBibitem
\bibitem[Fontcuberta~i Morral et~al.(2008)Fontcuberta~i Morral, Spirkoska,
  Arbiol, Heigoldt, Morante, and Abstreiter]{Morral08}
Fontcuberta~i Morral,~A.; Spirkoska,~D.; Arbiol,~J.; Heigoldt,~M.;
  Morante,~J.~R.; Abstreiter,~G. \emph{Small} \textbf{2008}, \emph{4},
  899--903\relax
\mciteBstWouldAddEndPuncttrue \mciteSetBstMidEndSepPunct{\mcitedefaultmidpunct}
{\mcitedefaultendpunct}{\mcitedefaultseppunct}\relax \EndOfBibitem
\bibitem[Krogstrup et~al.(2010)Krogstrup, Popovitz-Biro, Johnson, Madsen,
  Nygard, and Shtrikman]{Krogstrup10}
Krogstrup,~P.; Popovitz-Biro,~R.; Johnson,~E.; Madsen,~M.~H.; Nygard,~J.;
  Shtrikman,~H. \emph{Nano Letters} \textbf{2010}, \emph{10}, 4475--4482\relax
\mciteBstWouldAddEndPuncttrue \mciteSetBstMidEndSepPunct{\mcitedefaultmidpunct}
{\mcitedefaultendpunct}{\mcitedefaultseppunct}\relax \EndOfBibitem
\bibitem[Dhaka et~al.(2012)Dhaka, Haggren, Jussila, Jiang, Kauppinen, Huhtio,
  Sopanen, and Lipsanen]{Dhaka12}
Dhaka,~V.; Haggren,~T.; Jussila,~H.; Jiang,~H.; Kauppinen,~E.; Huhtio,~T.;
  Sopanen,~M.; Lipsanen,~H. \emph{Nano Letters} \textbf{2012}, \emph{12},
  1912--1918\relax
\mciteBstWouldAddEndPuncttrue \mciteSetBstMidEndSepPunct{\mcitedefaultmidpunct}
{\mcitedefaultendpunct}{\mcitedefaultseppunct}\relax \EndOfBibitem
\bibitem[Ruda and Shik(2005)Ruda, and Shik]{Ruda05}
Ruda,~H.~E.; Shik,~A. \emph{Phys. Rev. B} \textbf{2005}, \emph{72},
  115308\relax
\mciteBstWouldAddEndPuncttrue \mciteSetBstMidEndSepPunct{\mcitedefaultmidpunct}
{\mcitedefaultendpunct}{\mcitedefaultseppunct}\relax \EndOfBibitem
\bibitem[Wang et~al.(2001)Wang, Gudiksen, Duan, Cui, and Lieber]{Wang01}
Wang,~J.; Gudiksen,~M.~S.; Duan,~X.; Cui,~Y.; Lieber,~C.~M. \emph{Science}
  \textbf{2001}, \emph{293}, 1455--1457\relax
\mciteBstWouldAddEndPuncttrue \mciteSetBstMidEndSepPunct{\mcitedefaultmidpunct}
{\mcitedefaultendpunct}{\mcitedefaultseppunct}\relax \EndOfBibitem
\bibitem[Mishra et~al.(2007)Mishra, Titova, Hoang, Jackson, Smith,
  Yarrison-Rice, Kim, Joyce, Gao, Tan, and Jagadish]{Mishra07}
Mishra,~A.; Titova,~L.~V.; Hoang,~T.~B.; Jackson,~H.~E.; Smith,~L.~M.;
  Yarrison-Rice,~J.~M.; Kim,~Y.; Joyce,~H.~J.; Gao,~Q.; Tan,~H.~H.;
  Jagadish,~C. \emph{Applied Physics Letters} \textbf{2007}, \emph{91},
  263104\relax
\mciteBstWouldAddEndPuncttrue \mciteSetBstMidEndSepPunct{\mcitedefaultmidpunct}
{\mcitedefaultendpunct}{\mcitedefaultseppunct}\relax \EndOfBibitem
\bibitem[Birman(1959)]{Birman59}
Birman,~J.~L. \emph{Phys. Rev. Lett.} \textbf{1959}, \emph{2}, 157--159\relax \mciteBstWouldAddEndPuncttrue
\mciteSetBstMidEndSepPunct{\mcitedefaultmidpunct} {\mcitedefaultendpunct}{\mcitedefaultseppunct}\relax \EndOfBibitem
\bibitem[Dif()]{DifferentialPL12}
A moving window average over 10-50 pixels is subtracted from each spectrum.
  This produces a result which is almost indistinguishable from the numerical
  second derivative and serves to highlight small features (peaks) on a large
  background.\relax
\mciteBstWouldAddEndPunctfalse \mciteSetBstMidEndSepPunct{\mcitedefaultmidpunct} {}{\mcitedefaultseppunct}\relax
\EndOfBibitem
\end{mcitethebibliography}

\providecommand*\mcitethebibliography{\thebibliography} \csname @ifundefined\endcsname{endmcitethebibliography}
  {\let\endmcitethebibliography\endthebibliography}{}

\end{document}